\documentclass[12pt]{book}
\usepackage{graphicx}
\usepackage{pazh}
\usepackage{lscape}
\tightenlines
\begin{document}

Astronomy Letters, Vol. 31, No. 10, 2005, pp. 681 694. Translated from Pis'ma v Astronomicheskii Zhurnal, 
Vol. 31, No. 10, 2005, pp. 764 779. Original Russian Text Copyright $\copyright$ 2005 by Chelovekov, Lutovinov, Grebenev, Sunyaev.

\title
{\bf Observations of the X-ray Burster MX 0836-42 by the INTEGRAL and RXTE Orbiting Observatories}

\author
{\bf I. V. Chelovekov\affilmark{1*}, A.A. Lutovinov\affilmark{1}, S.A. Grebenev\affilmark{1}, R.A. Sunyaev\affilmark{1,2}}

\affil{{\it $^1$ Space Research Institute, Russian Academy of Sciences, Profsoyuznaya ul. 84/32, Moscow, 117810 Russia}\\ 
{\it $^2$ Max-Planck-Institut fur Astrophysik, Karl-Schwarzschild-Str. 1, Postfach 1317, D-85741 Garching, Germany}\\ }

\vspace{3mm}

\def\deg{^\circ}

\vspace{5mm}

\subsection*{}
{\bf Abstract.}
We present the results of our study of the emission from the transient burster MX 0836-42 using its 
observations by the INTEGRAL and RXTE X-ray and gamma-ray observatories in the period 2003-2004. The 
source's broadband X-ray spectrum in the energy range 3-120 keV has been obtained and investigated 
for the first time. We have detected 39 X-ray bursts from this source. Their analysis shows that the 
maximum 3-20 keV flux varies significantly from burst to burst, $F \sim (0.5-1.5) \times 10^{-8} ~erg ~cm^{-2} ~s^{-1}$. 
Using the flux at the maximum of the brightest detected burst, we determined an upper limit for the 
distance to the source, $D\simeq 8$ kpc.

Key words: neutron stars -- bursters, transients; X-ray sources -- \mbox{MX 0836-42}.

\vspace{8cm}
$^{*}$ e-mail: chelovekov@hea.iki.rssi.ru

\section*{ INTRODUCTION }

The transient X-ray source \mbox{MX 0836-42} was discovered in 1971 by the OSO-7 satellite 
(Markert et al. 1975). Its position almost coincided with the probable position of 
the point source detected in December 1970 and February 1971 by the UHURU observatory 
(Kellogg et al. 1971). Since the intensity of the latter was too low, it was not 
included in the official UHURU catalog of sources (Markert et al. 1977; Cominsky et al. 1978).
In 1990, the WATCH all-sky X-ray monitor aboard the GRANAT orbiting observatory detected 
a bright transient source near \mbox{MX 0836-42} (Sunyaev et al. 1990, 1991; Lapshov et al. 1992). 
Its 5-15 keV flux during these observations reached a level comparable to the flux from 
the Crab Nebula. The localization accuracy of the source was only $1\deg$. Analysis of the 
ROSAT data revealed two point sources in this region spaced 24 arcmin apart (Hasinger et al. 1990). 
Subsequently, the presence of these two sources was confirmed by data from the ART-P 
telescope of the GRANAT observatory (Sunyaev 1991). Type-I X-ray bursts were detected 
from the northern source, which allows \mbox{MX 0836-42} to be classified as a low-mass Xray 
binary containing a neutron star with a weak magnetic field, while X-ray pulsations with 
a period of $\sim12$ s were detected from the southern source GRS 0834-430, which characterize 
it as an X-ray pulsar (Makino 1990; Grebenev and Sunyaev 1991). Studies of the emission 
from \mbox{MX 0836-42} showed that its energy spectrum could be described by a power law with a 
photon index of $\sim1.5$ (Aoki et al. 1992). In this paper, based on the data obtained in 
2003-2004 by the instruments of the INTEGRAL and RXTE orbiting observatories, we have 
constructed and analyzed the source's spectrum during X-ray bursts and for the first time 
persistent broadband spectrum in the energy range 3-120 keV. We discuss the properties 
of the detected X-ray bursts.

\section*{\bf OBSERVATIONS AND DATA ANALYSIS }

\vspace{-3mm}

The INTEGRAL international orbiting gamma-ray observatory (Winkler et al. 2003) was placed 
in orbit by a Russian PROTON launcher on October 17, 2002 (Eismont et al. 2003). There are 
four instruments aboard the observatory: the SPI gamma-ray spectrometer, the IBIS gamma-ray 
telescope, the JEM-X X-ray monitor, and the OMC optical monitor. Here, we use the data 
obtained by the ISGRI detector, one (upper) of the two detectors of the IBIS gamma-ray 
telescope (Ubertini et al. 2003), and by the second module of the JEM-X X-ray monitor 
(Lund et al. 2003). The ISGRI/IBIS detector is sensitive to photons in the energy range 
15-200 keV and has an energy resolution of $\sim 7\%$ at 100 keV. The IBIS telescope includes a 
coded mask that allows it to be used not only for spectral and timing analyses of the 
emission, but also for reconstructing the image of the sky in the $29 \deg \times 29 \deg$ field of view 
of the instrument (the fully coded field of view is $9\deg \times 9\deg$) with an angular resolution 
of 12 arcmin (FWHM) and localizing X-ray and gamma-ray sources to within 1-2 arcmin. The 
JEM-X telescope is also based on the principle of a coded aperture. It is sensitive to 
photons in the energy range 3-35 keV, and its fully coded field of view is $4\deg .8$ in diameter. 

We analyzed the JEM-X and IBIS data using the OSA 4.1 data processing software package 
distributed by the INTEGRAL Science Data Center (ISDC). To construct the light curves and 
spectra for \mbox{MX 0836-42}, we used the fluxes that were obtained by reconstructing the image 
of the sky in the field of view of the instrument and identifying the observed sources. The 
photon spectrum of the source in 20-120 keV energy band was reconstructed using a 50-channel 
ISGRI/IBIS response matrix that was constructed from observations of the source in the Crab Nebula and that 
allows us to restore the spectral shape of the source to within 4\% and the normalization to 
within 7\%. 

The RXTE observatory carries three main instruments: the PCA spectrometer based 
on five xenon proportional counters sensitive to photons in the energy range 2-60 keV, the 
HEXTE spectrometer sensitive to photons up to 200 keV, and the ASM all-sky monitor sensitive 
to photons in the energy range 2-12 keV. The RXTE observational data for the source under study 
were provided by the NASA archive (HEASARC). We used the LHEASOFT 5.3.1 software package and 
the XSPEC 11.3.1 code to process the PCA/RXTE and HEXTE/RXTE data and to analyze the source's 
spectra. 

The X-ray transient \mbox{MX 0836-42} was within the IBIS/ISGRI field of view several times in 
the period from March 2003 through May 2004 (Table 1), within the framework of both the Core and 
Open observing programs (Winkler et al. 2003). The total exposure time for this source was more 
than 2.5 Ms. We used the data obtained when scanning the Galactic plane and during deep observations 
of a region near the source Vela X-1. 

The most recent accessable observations of the above-mentioned Vela X-1 
region were performed from June 12 through July 6 and from November 27 through December 11, 2003. 
The total JEM-X and IBIS exposure times for MX 0836-42 were $\sim0.73$ and $\sim1.13$ Ms for the former 
period and $\sim0.45$ and $\sim1.0$ Ms for the latter period, respectively. We used only the pointings 
during which the source under study was within the fully coded field of view of the instrument. 
This was done to avoid the inaccuracies in restoring the energy flux from the source under study 
as much as possible. The difference between the exposure times for the two instruments stems from
the fact that the fully coded field of view of the JEM-X monitor is smaller than that of the IBIS 
telescope (see above).

Figure 1 shows the light curves of the source for the period 52650-53400 MJD constructed from the 
data of the ISGRI/IBIS (20-60 keV) and JEM-X (3-20 keV) telescopes aboard the INTEGRAL observatory 
and the ASM (2-12 keV) and PCA (3-20 keV) instruments aboard the RXTE observatory. Figure 2 presents 
the source's light curve constructed from all of the available ASM/RXTE data. The fluxes shown are 
the ratios of the fluxes from \mbox{MX 0836-42} to the flux from the Crab Nebula in the corresponding energy 
range. 

During two of the three groups of measurements consisting of nine (52773-52825 MJD) and six 
(53186-53261 MJD) observations, respectively, present in Fig. 1a, the flux from the source under 
study was below the sensitivity threshold of the ISGRI detector. The upper limits on the flux from 
\mbox{MX 0836-42} for each of these pointings are given at the $3\sigma$ level. 

Figures 1b and 2 show the light curve for \mbox{MX 0836-42} in the energy range 2-12 keV constructed from 
the ASM/RXTE data. Each point in the figure corresponds to the flux from the source averaged over 
a 36-ks period. 

Table 2 presents information about the observations by the RXTE orbiting observatory in 2003-2004 
during which \mbox{MX 0836-42} was within the PCA and HEXTE fields of view. Figure 3b shows the 3-120 keV 
fluxes from this source determined by processing the RXTE (PCA+HEXTE) data. Here, we use only the 
sessions of stable pointings of the instruments at the source.

\section*{\bf THE PERSISTENT SPECTRUM}

Based on the INTEGRAL data, we were able to construct the 3-120 keV spectrum of the source under study 
(Fig. 4, dashes). For this purpose, we used the observational data obtained by the JEM-X X-ray monitor 
(3-20 keV; Fig. 1c, region II) and the ISGRI/IBIS detector (20-120 keV; Fig. 1a, region II) during 137-141 
orbital cycles (Table 1, November 27 - December 9, 2003), when the source was in its high state and its flux 
in these energy ranges was $\sim 50-70$ mCrab. During the fitting, the JEM-X spectrum was renormalized to 
correspond to the normalization of the ISGRI/IBIS spectrum; the normalization factor was 1.15. We now 
attribute the spectral features in the regions 6-8, 12-15 and 20-25 keV to systematic measurement errors. 

A spectral analysis of the emission from \mbox{MX 0836-42} using the data averaged over each of the INTEGRAL orbital 
cycles mentioned above showed that the spectral shape of the source did not change significantly over this 
period. Therefore, we used the averaged data of these six orbital cycles (Fig. 1a, region II) to construct 
and analyze the broadband spectrum. Fitting the constructed spectrum of the source by a power law with a 
high energy exponential cutoff yields a photon index of $\alpha = 1.46 \pm 0.08$ and a cutoff energy of 
$E_{cut}=51.1\pm1.4$; the 3-120 keV flux from the source was $F=(2.29\pm0.10) \times 10^{-9} ~erg ~cm^{-2} ~s^{-1}$. 

Based on the ISGRI/IBIS data averaged over 55-58 
(Fig. 1a, region I) and 149-194 (Fig. 1a, region III) orbital cycles, we constructed the source's spectra 
only in the energy range 20-60 keV, since the total exposure times in these regions were only $\sim20$ and 
$\sim50$ ks, respectively, which are much shorter than the exposure time in region II ($\sim 1$ Ms). The statistical 
significance of the source's detection at energies above 60 keV during these observations was insuficient to 
construct a qualitative spectrum. Fitting these spectra by a power law with a high-energy exponential cutoff
yields a photon index $\alpha=1.29\pm0.17$ for the former and $\alpha=1.31\pm0.14$ for the latter ($E_{cut}$ 
during these fitting was fixed at 50 keV).

Figure 4 (solid lines) shows examples of the spectra for \mbox{MX 0836-42} that were obtained from the PCA (3-20 keV) 
and HEXTE (20-60 keV) data averaged over several successive pointings (the total exposure time is $\sim (2.9-10.5)$ ks). 
All of the spectra obtained were fitted in the energy range 3-60 keV by a power law with a high energy 
exponential cutoff. We also added the reflection of radiation from the accretion disk, the photoabsorption 
under the assumption of solar heavy-element abundances in the interstellar medium, and the fluorescence iron line 
at $E_{_{Fe}}=6.4$ keV to this model, which allowed the quality of the fit to be improved considerably (the $\chi ^2$
value per degree of freedom decreased from $\sim (8-10)$ to $\sim (1-2)$). In view of the uncertainty in the 
normalization of the HEXTE spectra, all of them were multiplied by a constant to be renormalized to the level 
of the PCA spectra obtained during the same observation. 

Table 3 presents the fitting results and the model fluxes 
corrected for the dead time of the detector and the HEXTE spectral normalization constants mentioned above. 
Figure 3 shows the time dependences of the photon index, the source's persistent model flux, and the hydrogen 
column density ($N_{H}$) derived when fitting the spectra. The dotted line in this figure highlights the 
parameters determined from the data of January $29_1$ and $31_2$ (the subscript indicates the number of a given 
pointing among the pointings at the source on this day) and February 2, 2003. The relatively high values of $N_{H}$ 
and the low values of the photon index and the flux obtained when fitting these data allow us to separate them
into a group of observations with strong absorption. The derived mean photon index of the source's power-law 
spectrum, $\sim (1.4-1.5)$, and the interstellar absorption, $\sim 3 \times 10^{22} ~atoms ~cm^{-2}$ (Fig. 3, 
without including the observations from the group with strong absorption mentioned above), are comparable 
to those obtained previously when studying this source (Aoki et al. 1992; Belloni et al. 1993). 

The iron emission line at $E_{_{Fe}}=6.4$ keV was detected in all PCA spectra. Since the PCA spectral resolution is 
too low for the line profile to be studied in detail, we fixed the line parameters at $E_{_{Fe}}=6.4$ keV 
and $\delta E_{_{Fe}}=0.1$ keV when fitting the spectra. The line equivalent width in the spectra under 
consideration was 100-310 eV.

\section*{\bf X-RAY BURSTS}

When analyzing the JEM-X 3-20 keV light curves for \mbox{MX 0836-42}, we found 24 X-ray bursts (Table 4). The light 
curves of this source were constructed only for the period of its reliable detection above the background level 
from the data obtained during 137-141 and 146 orbital cycles (Fig. 1, region III; Table 1, November 27 - 
December 24, 2003). 

There were no pointings containing more than one burst. The separation between the nearest of the neighboring
bursts was $\sim 2$ h, in good agreement with the burst recurrence period for this source estimated previously 
(Aoki et al. 1992). 

To carry out a detailed analysis of the bursts from the source under study in the energy range 3-20 keV, we 
used the 25 observations of \mbox{MX 0836-42} performed by the PCA detector of the RXTE orbiting observatory from 
January 24 through March 20, 2003, and from January 18 through January 26, 2004 (Table 2). We found 15 X-ray 
bursts in the light curves constructed from these data. The source's radiation temperature during the decay 
of these bursts decreased (Fig. 5b), which is characteristic of type I X-ray bursts (Lewin and Joss 1981). 
Analysis of the observations during which the observatory was repointed revealed no new burst.

The burst flux from the source reached its maximum, on average, in 6-8 s (Fig. 5b) and then remained at the 
same level for 3-4 s during some of the bursts (Fig. 5c). Table 5 gives the burst durations and the exponential 
burst decay times. The burst duration was defined as the ratio of the total energy released in the burst component 
of the emission from the burst onset time to the time the flux decreased to 10\% of its maximum to the mean burst 
energy flux over this period. To calculate the exponential burst decay time, we fitted the burst profile by an 
exponential time dependence of the flux. 

A characteristic feature of 80\% of the X-ray bursts detected by the PCA
spectrometer from \mbox{MX 0836-42} is a more or less distinct double-peaked structure. An example of such a burst is 
shown in Fig. 5a. It is believed that a multipeaked burst shape can result from the following: (i) expansion of 
the photosphere under the pressure of a near-Eddington flux and (ii) peculiarities of the thermonuclear burning. 
Since we found no statistically significant increase in the color radius of the emitting object at the time of the 
dip between the peaks (the 3-20 keV flux decreased by $\sim 15\%$), we can assume that the double-peaked structure 
of the burst in this case is not related to photospheric expansion, but could result from peculiarities of the 
thermonuclear burning in the source during the burst. 

Interestingly, the maximum 3-20 keV flux from the source 
during the X-ray bursts detected by the PCA spectrometer ranged from $F \sim 1.5 \times 10^{-8} ~erg ~cm^{-2} ~s^{-1}$
(Fig. 5c) to $F \sim 0.5 \times 10^{-8} ~erg ~cm^{-2} ~s^{-1}$ (Fig. 5d). This allows us to trace the dependence 
of this quantity on the total energy released during the burst (Fig. 6a). We see that the maximumburst flux rises 
with increasing total energy released during the burst. A similar dependence was found for other sources of X-ray 
bursts, such as 1608-522 (Murakami et al. 1980), 1728-337 (Basinska et al. 1984), 1735-44 (Lewin et al. 1980), 
and 1837+049 (Sztajno et al. 1983). 

Since no photospheric expansion of the neutron star was reliably detected in 
any of the bursts studied, we used the flux at the maximum of the brightest of the X-ray bursts mentioned above 
($1.12\pm0.24$ Crab) to estimate the distance to \mbox{MX 0836-42} by assuming that the source's luminosity at this time 
was close to the Eddington limit for a neutron star with a mass of 1.4 $M_{\sun}$. The derived upper limit for the distance 
to the source is $D\sim 8$ kpc. This value is close to the lower limit for the distance to the source $D \sim (10-20)$ kpc 
estimated previously by Aoki et al. (1992) by assuming that the color radius of the emitting object during an X-ray burst 
would correspond to the neutron-star radius, 10 km. 

Figures 5c and 5d show the time dependences of the model flux 
and the color temperature and radius of the emitting object obtained when fitting the source's spectra by a blackbody 
during the brightest and weakest bursts detected by the PCA/RXTE spectrometer. All of the spectra studied were corrected 
for the background count rate of the detector and the persistent emission from the source under study. The color 
temperature of the emitting region, on average, rose at the burst onset to 2-2.5 keV in 1-4 s and gradually fell to 
1.5-2.0 keV during the burst. The color radius $R_{c}$ of the emitting region, on average, rose at the burst onset 
from 1-3 to 4-6 km in 3-5 s and decreased insignificantly by the burst end (by $\sim 10-15 \%$). This behavior of $R_{c}$ 
may suggest that the size of the region affected by the explosion changes and that Comptonization plays a prominent 
role in shaping the spectrum. The mean values of the maximum temperature, $kT_{_{bb}} \sim 2.5$ keV, and the radius, 
$R\sim (4-6) \times$ (D/8 kpc) km, are in good agreement with the values of these parameters obtained by Aoki et al. (1992). 

Figure 6b shows the dependence of the persistent flux from the source on the total energy released during the burst 
constructed from the 15 X-ray bursts detected by the PCA/RXTE spectrometer. We see from this figure that there is a 
direct correlation between these quantities. This can serve as direct evidence for the current understanding of the 
burster phenomenon. Accreting matter falls to the neutron star surface in the time between bursts, releasing part of 
its gravitational energy in the form of radiation that we observe as the system's persistent emission. Subsequently, 
this matter becomes a fuel for stable and explosive thermonuclear reactions, the latter of which are observed as an 
X-ray burst. In this case, if we assume that the entire accumulated store of fuel is used up during a burst, then 
the total energy released during the burst increases with persistent flux from the system and, hence, with accretion 
rate. 

Since the PCA spectrometer is not a telescope, i.e., the sky in the field of view of the instrument cannot be 
imaged, we cannot assert with confidence that precisely \mbox{MX 0836-42} is the source of the detected bursts. Note, however, 
that only one known burster, \mbox{MX 0836-42}, was within the PCA field of view when each of the bursts was detected. 

The RXTE observatory is in a low near-Earth orbit; therefore, its instruments can continuously monitor the source only during 65\% 
of its 90-min orbit. Since more than one X-ray burst occurred in none of the PCA observing sessions that we used, we cannot 
reliably determine the burst recurrence time $\bf{\tau_{_R}}$ for \mbox{MX 0836-42} from these data. However, in the two successive 
sessions on January 31, 2003, the PCA spectrometer detected two X-ray bursts separated by an interval of $\sim$ 7205 s, 
which corresponds to the burst recurrence period determined for this source by Aoki et al. (1992). The source was 
continuously observed for $\sim$ 2850 s after the first of these bursts; subsequently, the observations were interrupted 
for $\sim$ 2610 s and then resumed. The 3-20 keV flux averaged over the second of the bursts was $F=(3.50\pm0.81) 
\times 10^{-9} ~erg ~cm^{-2} ~s^{-1}$, and the mean integrated flux from the source in the energy range 3-120 keV between 
these bursts was $F=(1.61\pm0.06) \times 10^{-9} ~erg ~cm^{-2} ~s^{-1}$. We assume that the bulk of the radiative energy 
during the burst is released in the range 3-20 keV. 

In accordance with the current understanding of the burster phenomenon, 
the gravitational energy $E_{_g}$ of the matter accreted by the neutron star is released between X-ray bursts from such 
a system and the energy $E_{_b}$ of its thermonuclear burning is released during bursts. If we assume that no other bursts 
occurred over the interval between the pointings containing the first and second bursts during which the system was not 
monitored, then we can determine the thermonuclear burning parameter, the ratio $\alpha = \frac{E_{_g}}{E_{_b}}$, from 
the relationship

\centerline{\hspace{5.5cm} $\alpha \approx \tau_{_R} L_{_P} / \tau_{_B} L_{_B}$, \hspace{5.5cm} (1)}

where $L_{_P}=1.24 \times 10^{37} (D/8 kpc)^2 ~erg ~s^{-1}$ and $L_{_B}=5.04 \times 10^{37} (D/8 kpc)^2 ~erg ~s^{-1}$
are the mean persistent and burst luminosities of the system and $\tau_{_B}=12.3$ s and $\tau_{_R}=7205$ s are the second 
burst duration and the period between bursts, respectively. In our case, formula (1) yields $\alpha \approx 144$, typical 
of a burst occurring through helium detonation (Bildsten 2000). 

If we assume that the regime of thermonuclear burning did 
not change from burst to burst, then we can estimate the burst recurrence periods in other sessions using the above value 
of $\alpha$. Table 5 gives the burst recurrence times for the source under study determined using formula (1) by assuming
that $\alpha \approx 140$. Comparison between the recurrence periods and the mean duration of the source's continuous 
monitoring by the observatory lets the fact of observation of only one burst in each of the sessions can be easily explained. 
It is worth mentioning that the estimates obtained by this method are a factor of 2-3 larger than the burst recurrent period 
for \mbox{MX 0836-42} determined by Aoki et al. (1992): $\tau_{_R} \sim 2$ h. If we assume that $\tau_{_R} \sim 2$ h, then we can 
determine the parameter $\alpha \sim 80$ averaged over all of the observed bursts using formula (1); this value is typical of 
mixed hydrogen helium bursts (Bildsten 2000). This model describes better the shape of the observed bursts, in particular, 
the relatively long (6-8 s) period of the burst rise to its maximum level. 

Table 2 gives the accretion rate corresponding 
to the detected persistent integrated flux from the source under study and the corresponding recurrence periods of hydrogen 
helium bursts from it calculated by assuming that accreted matter occupies 1/3 of the surface ($\tau_{1}$) and the entire 
surface ($\tau_{2}$) of the neutron star. We see from this table that if the bursts in \mbox{MX 0836-42} are hydrogen helium ones, 
then the accreted matter involved in the explosion during the burst occupies only part of the neutron star surface.

\section*{\bf DISCUSSION}

Figure 2 shows the light curve for \mbox{MX 0836-42} constructed from all of the available ASM data, in which we clearly 
see a rise in the flux from the source on a time scale of $\sim 600$ days. Clearly, even if this phenomenon is periodic, 
the period of such bursts is more than 7 years. For example, nonuniformity in the accretion process can be responsible 
for such bursts. 

It follows from Fig. 3 that there is a group of three observations with an anomalously high absorption 
level among the RXTE observations of the source (Table 3). It can be assumed that additional absorption possibly associated 
with the outer regions of the accretion disk appears in the system during these observations. The following fact argues for 
this explanation: the harder (more absorbed) the spectrum, the stronger the iron emission line in it. For example, the line 
equivalent width in the spectrum constructed from the observations on January 26 was $\sim 120$ eV, while this parameter 
for the $29_1$ observations was $\sim 310$ eV. This explanation suggests a large inclination of the accretion disk in the 
system under study. This process can be periodic and related to the orbital motion in the binary system; i.e., the source 
can be a dipper. To test this assumption, we analyzed the light curve of the source in the energy range 1.3-3.0 keV, which 
is subject to the strongest absorption, constructed from the ASM/RXTE data for the period from January 5, 1996, through 
April 7, 2005, for the presence of periodic variations in the frequency range $(5-300) \times 10^{-6}$ Hz, which is typical 
of the orbital motion of a low-mass binary. This analysis failed to reveal any significant period of the signal variations. 

The transient X-ray pulsar GRS 0834-43 is in the immediate vicinity of the source under study (24 arcmin). Since the PCA and 
HEXTE spectrometers are incapable of spatially resolving the sources of the detected emission, the above pulsar could introduce 
distortions in the observed spectrum of \mbox{MX 0836-42} when falling within the fields of view of these instruments. To test this 
hypothesis, we searched for the pulsations with a period of $\sim 12$ s that are typical of the pulsar GRS 0834-43 using all 
the available PCA observations of \mbox{MX 0836-42}. These studies failed to reveal any pulsating component whose confidence level 
would exceed 3 $\sigma$ in any of the observations. It is worth mentioning that the ASM/RXTE data revealed no flare activity 
in the emission from GRS 0834-43 during the period under consideration, while according to the ISGRI/IBIS data, the confidence 
level of the source's detection in the range 20-60 keV from January 2003 through March 2004 did not exceed 3 $\sigma$; the upper 
limit on the flux from the source was 1 mCrab throughout this period and $\sim 10$ mCrab for the period of a single observation 
($\sim 2$ ks). Thus, we can state the persistent spectra constructed from the available PCA and HEXTE data actually pertain 
to \mbox{MX 0836-42}. 

In the course of some of the 15 X-ray bursts detected by the PCA spectrometer during the pointings at the source under study, 
the mean color radius of the emitting object was 4-7 km (Figs. 5c and 5d), which is smaller than the value obtained in terms 
of the standard models for the structure of a neutron star for its radius. A modification of the source's spectrum through 
scattering, which leads to an overestimation of the color temperature and an underestimation of the color radius 
(London et al. 1986; Sunyaev and Titarchuk 1986; Babul and Paczynski 1987), could be responsible for this discrepancy. 

We see from Table 5 that the total energy released during the burst on January $31_2$, 2003, is a factor of 2-3 lower than 
that for the remaining bursts, and that the burst itself occurred a factor of 2 earlier than the accumulation of a matter 
column density enough for detonation could occur (if detonation affects the entire surface of the neutron star) (Table 2). 
A similar situation was observed in the emission from this source on February 18, 1991 (Aoki et al. 1992), except that the 
separation between the bursts in that case was only $\sim 10$ min. These phenomena can be explained as follows:

(1) Another burster is the burst source. Although this burst is morphologically similar to other bursts from the source 
under consideration, given that the PCA spectrometer is incapable of imaging the observed sky region, this assumption cannot 
be refuted using the available data. This assumption can also explain the different maximum burst energy fluxes from the source.

(2) The thermonuclear burning during the burst proceeds in a regime different from the remaining cases. This is possible if, 
for example, some amount of fuel was not used up during a previous burst. In cases 1 and 2, we can say nothing about the 
parameter $\alpha$ and the burst recurrent period in other sessions. 

(3) Accretion affected a smaller part of the neutron star surface, which led to a faster accumulation of matter to the column 
density required for detonation. This could be evidenced by the larger increase in the color radius at the burst onset and its 
larger decrease at the burst end than those for other bursts (Fig. 5d).

\section*{\bf ACKNOWLEDGMENTS}

We thank M.G. Revnivtsev for valuable suggestions and remarks made by him when preparing this paper. This work is based on 
the observational data obtained by the INTEGRAL observatory and provided through the Russian and European INTEGRAL Science 
Data Centers as well as the data obtained by the RXTE observatory and provided through the NASA's HEASARC Web page on the 
Internet (http://legacy.gsfc.nasa.gov). This study was supported by the Russian Foundation for Basic Research (project no. 
05-02-17454), the Presidium of the Russian Academy of Sciences (the  Nonstationary Astronomical Phenomena Program), and the 
Program for Support of Leading Scientific Schools (project no. NSh-2083.2003.2).

\begin{figure}[hp]
\centerline{\includegraphics[width=0.9\linewidth]{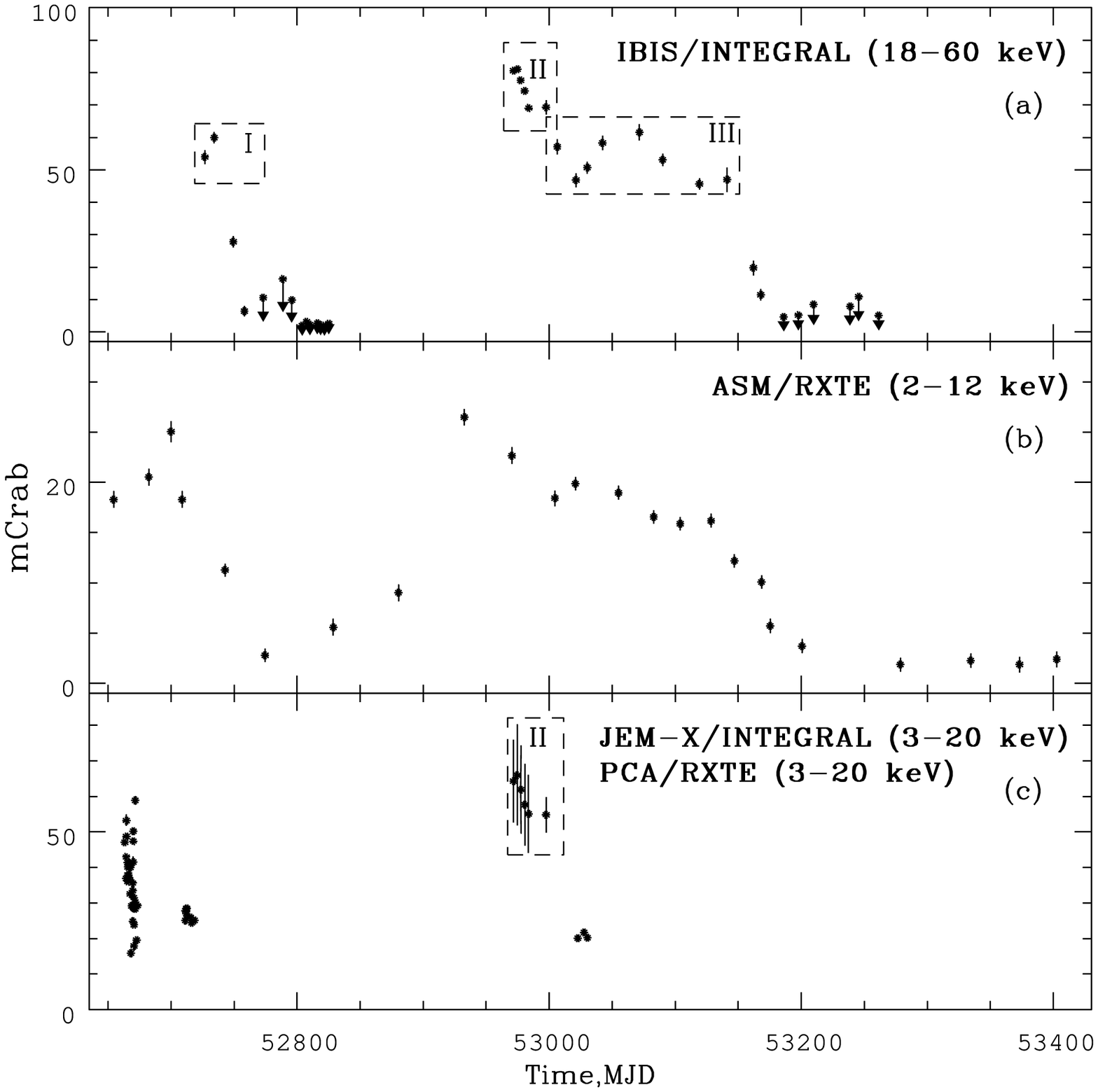}}
\caption{Light curves for MX 0836-42 constructed from (a) ISGRI/IBIS/INTEGRAL data, (b) ASM/RXTE data, and (c) PCA/RXTE and 
JEM-X/INTEGRAL data (region II). The upper limits on the flux are given at the $3 \sigma$ level.}
\end{figure}

\begin{figure}[hp]
\centerline{\includegraphics[width=0.9\linewidth]{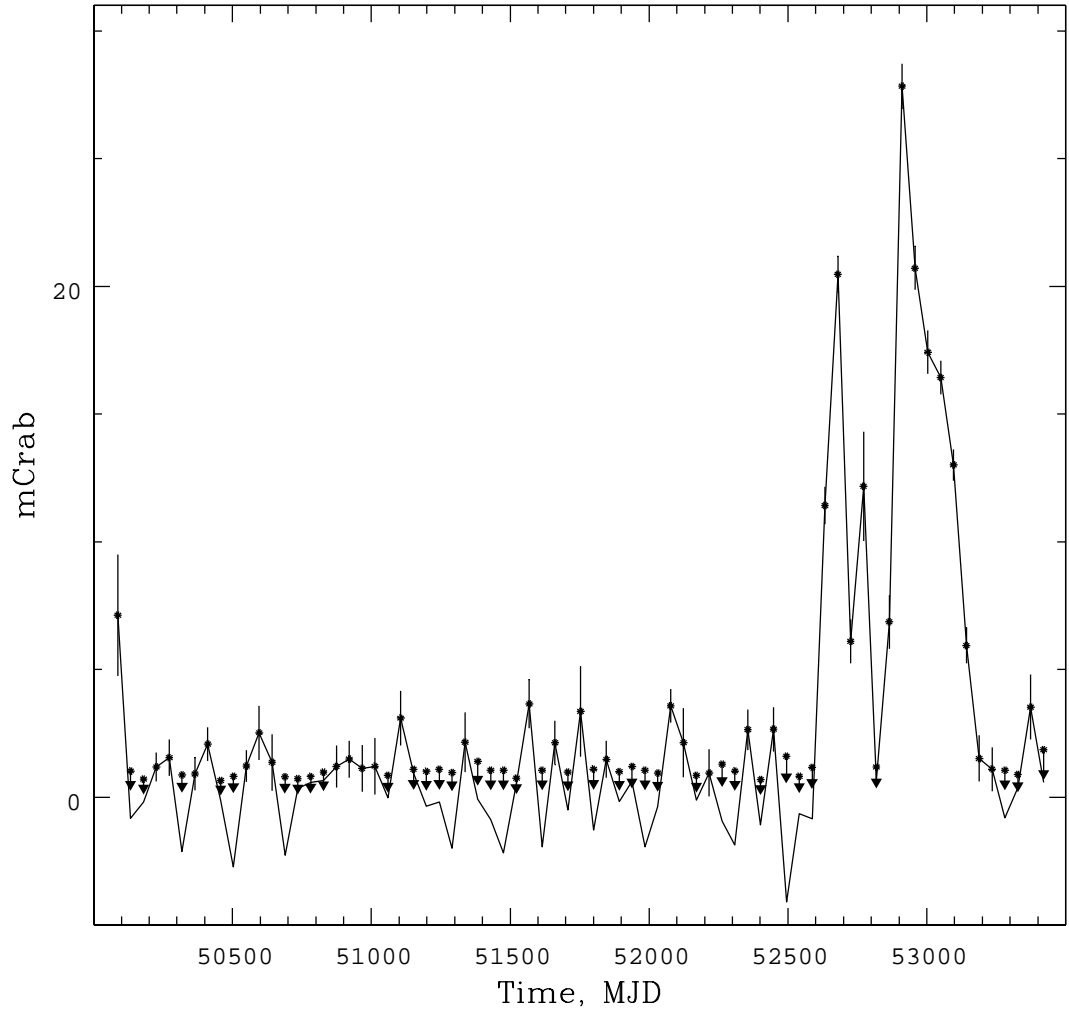}}
\caption{Light curve for MX 0836-42 in the energy range 2-12 keV constructed from ASM/RXTE data.}
\end{figure}

\begin{figure}[hp]
\centerline{\includegraphics[width=0.9\linewidth]{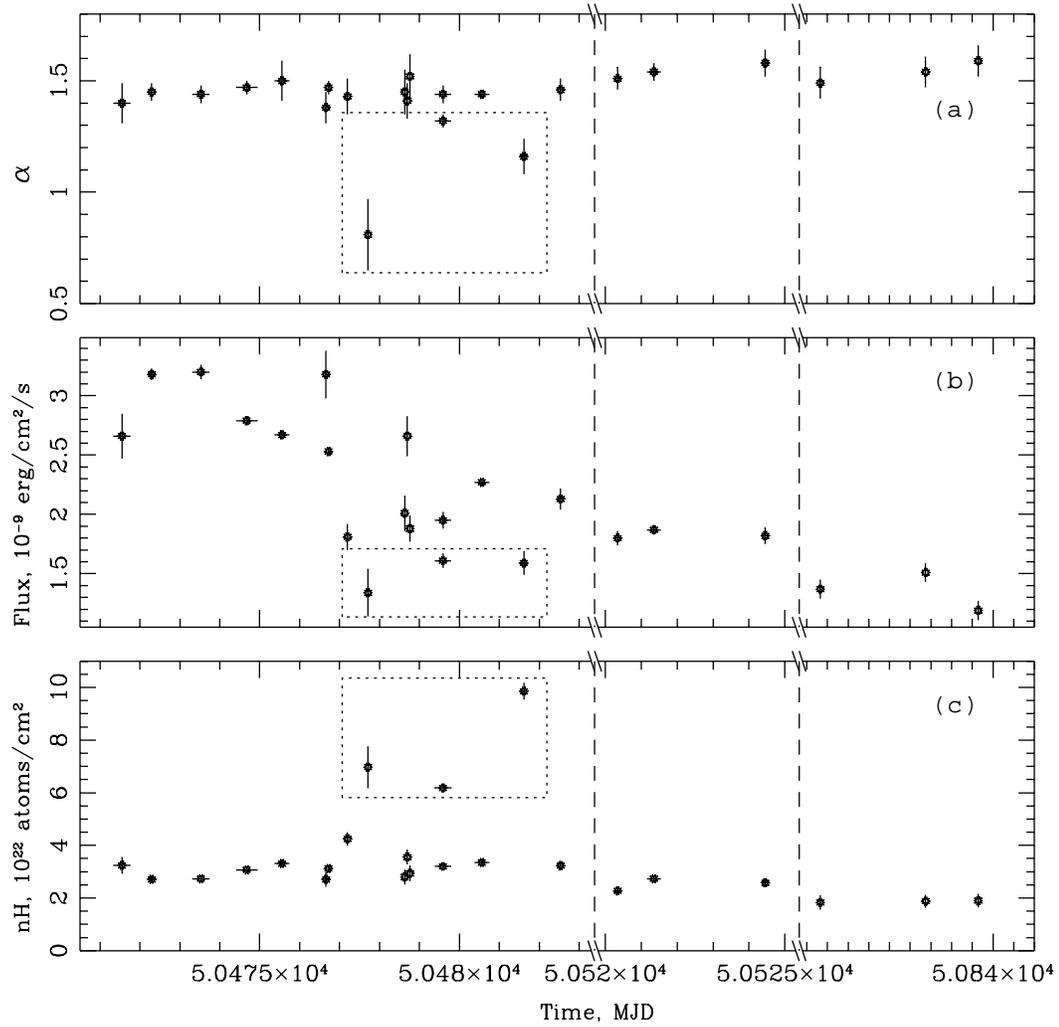}}
\caption{(a) Time dependences of the photon index, (b) the 3-120 keV flux, and (c) the interstellar absorption derived 
from the RXTE (PCA+HEXTE) observations of MX 0836-42 in 2003-2004.}
\end{figure}

\begin{figure}[hp]
\centerline{\includegraphics[width=0.9\linewidth]{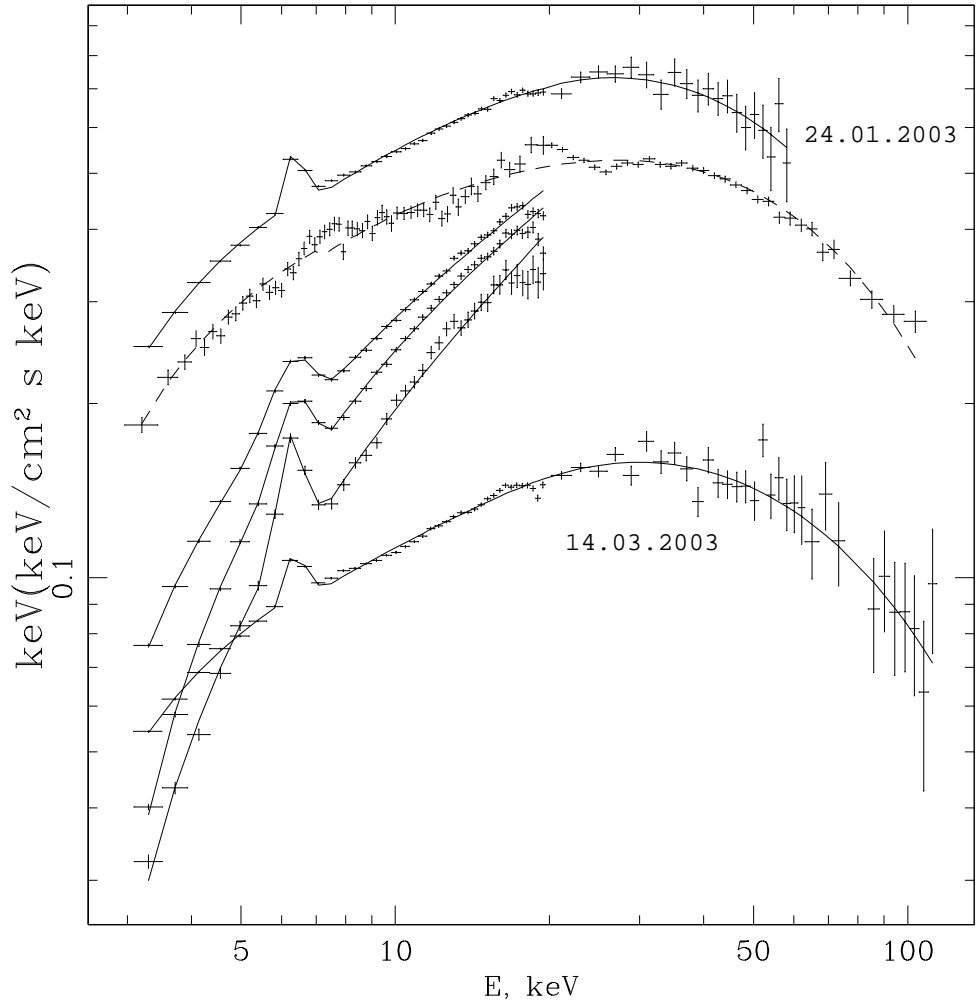}}
\caption{Persistent spectra for MX 0836-42 constructed from the data of the JEM-X and ISGRI/IBIS telescopes 
aboard the INTEGRAL observatory (dashed line) and the PCA (3-20 keV) and HEXTE (20-60 keV) detectors aboard 
the RXTE observatory (solid lines). For convenience of perception, the values of these spectra and the model 
in the spectrum for March 14, 2003, were multiplied by a factor of 1/3.}
\end{figure}

\begin{figure}[hp]
\centerline{
\includegraphics[width=0.45\linewidth]{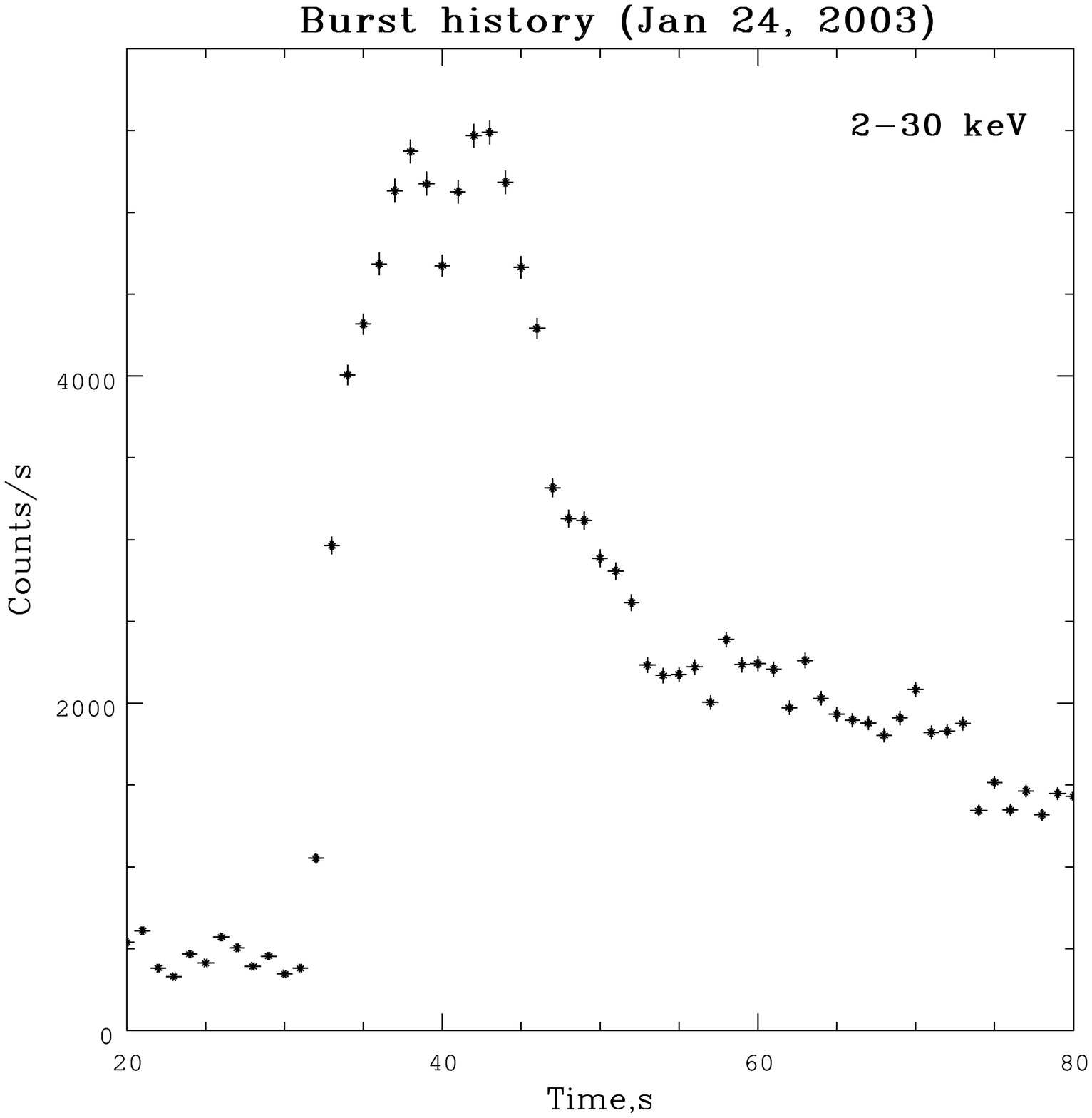}
\hspace{-.5cm}
\includegraphics[width=0.45\linewidth]{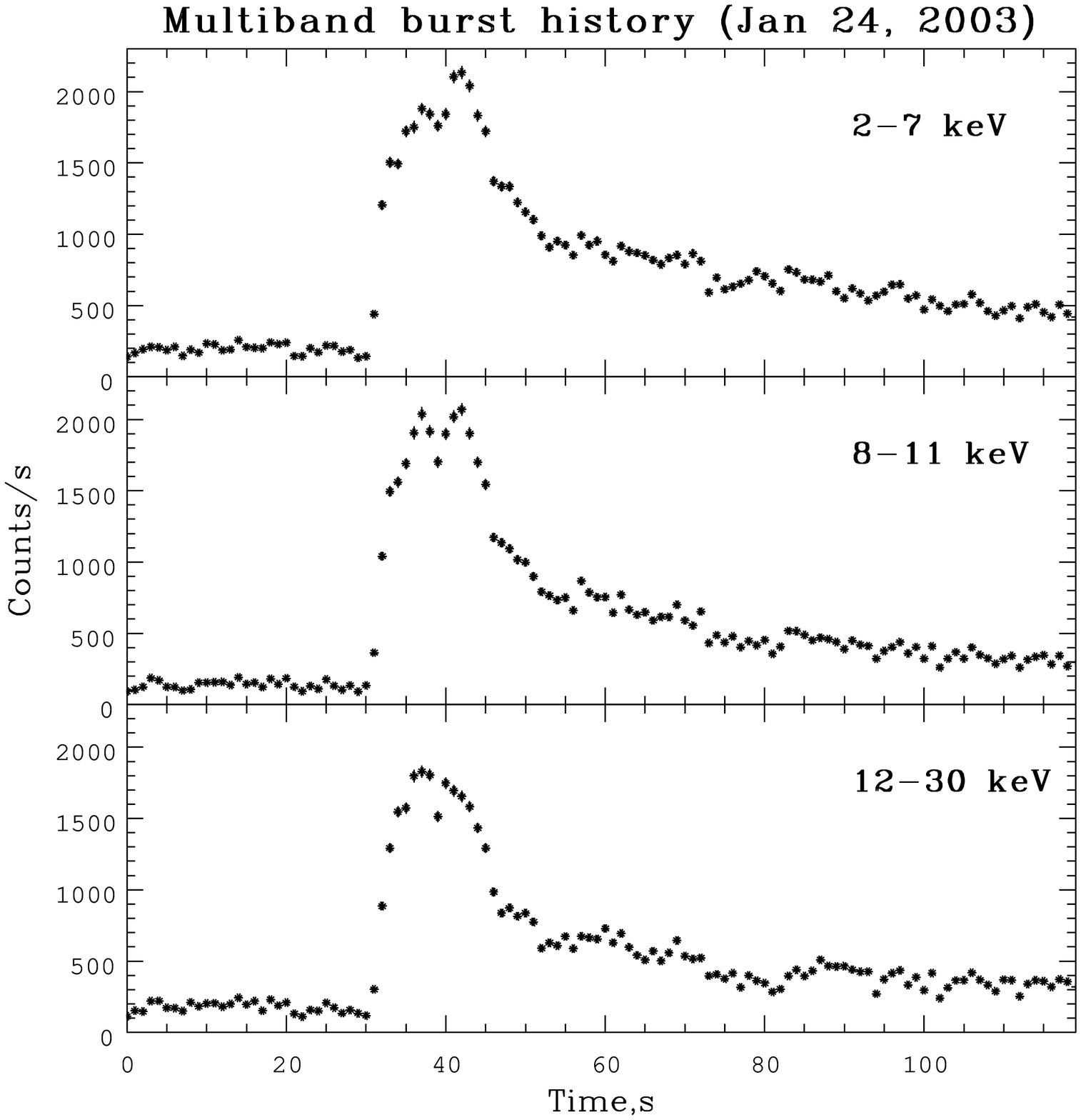}
}
\centerline{a~~~~~~~~~~~~~~~~~~~~~~~~~~~~~~~~~~~~~~~~~~~~~~~~~~~~~~b}
\centerline{
\includegraphics[width=0.45\linewidth]{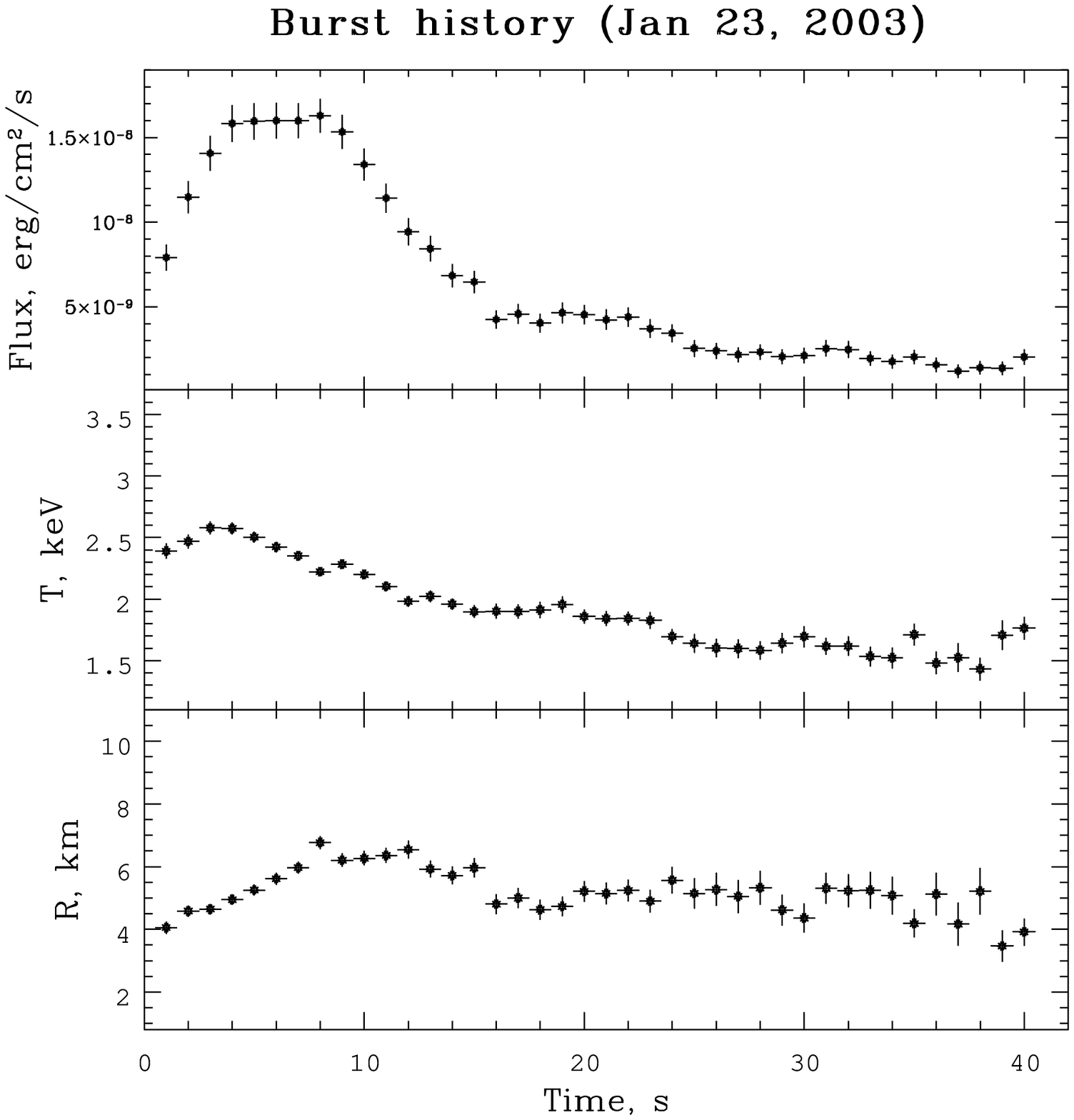}
\hspace{-.5cm}
\includegraphics[width=0.45\linewidth]{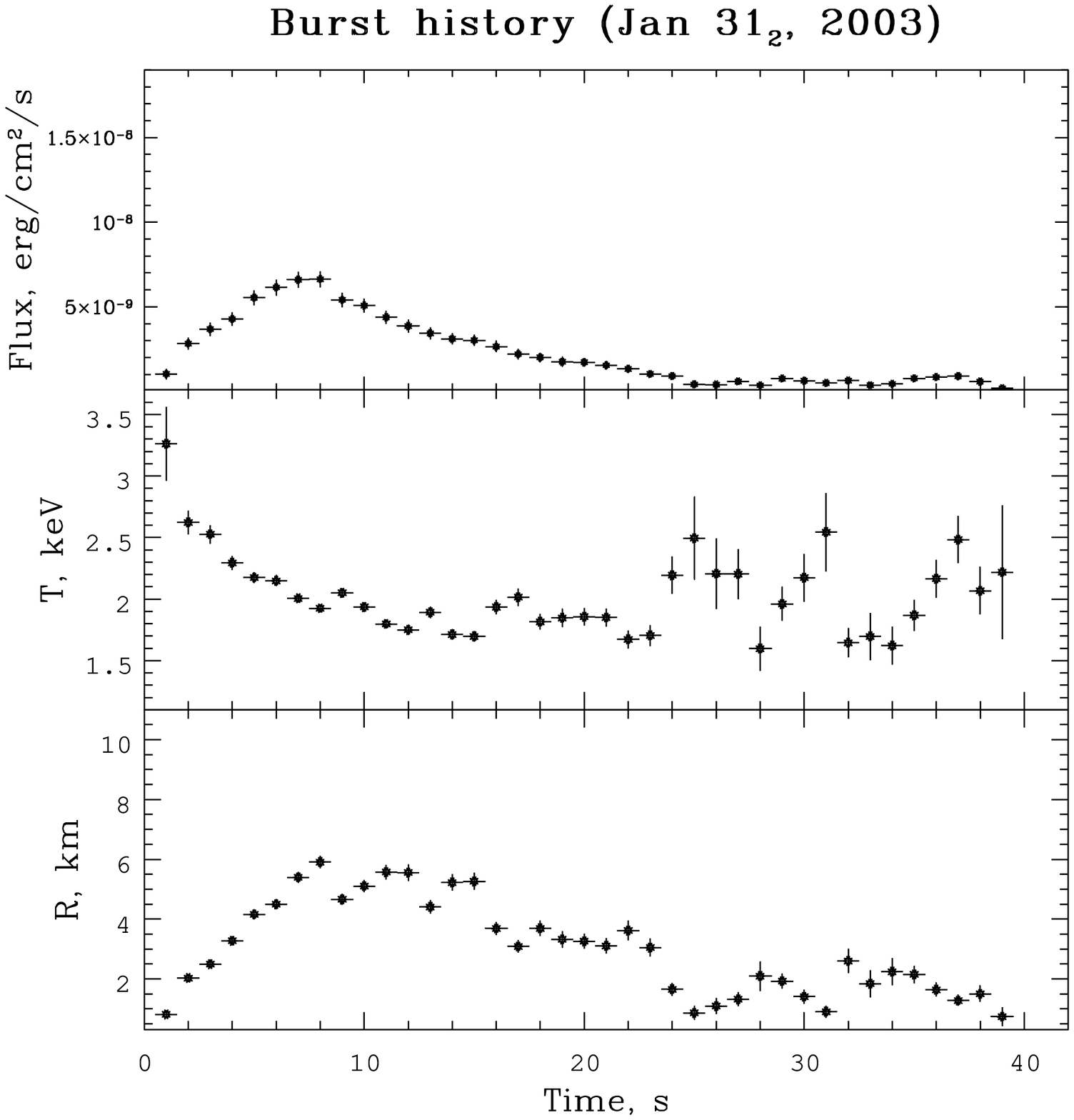}
}
\centerline{c~~~~~~~~~~~~~~~~~~~~~~~~~~~~~~~~~~~~~~~~~~~~~~~~~~~~~~d}
\caption{Panels (a) and (b) show the time histories of the burst detected on January 24, 2003, from MX 0835-42 
by the PCA/RXTE spectrometer in various energy ranges. A double-peaked structure of the burst near the maximum 
is clearly seen. Panels (c) and (d) display the time dependences of the 3-20 keV flux and the temperature and 
radius of the emitting object determined by fitting the spectra of MX 0836-42 by a blackbody during the bursts 
detected in 2003-2004 by PCA/RXTE. The time in panels (c) and (d) is measured from the burst onset; in panels 
(a) and (b), the burst onset corresponds to 30 s on the time axis.}
\end{figure}

\begin{figure}[hp]
\centerline{\includegraphics[width=0.9\linewidth]{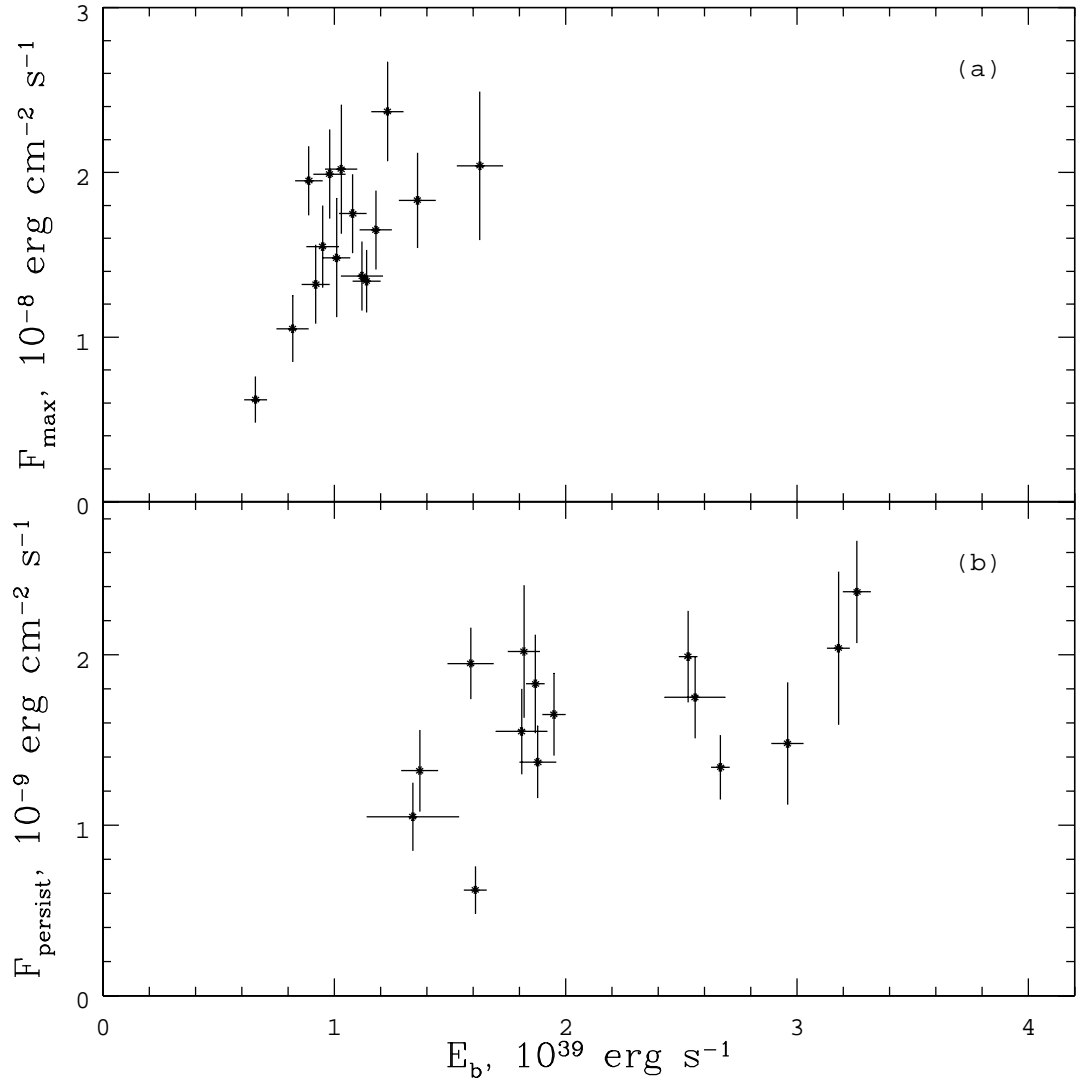}}
\caption{(a) Maximum burst flux versus total energy released during the burst and (b) persistent flux from the 
source vs. total energy released during the burst. The dependences were constructed for the 15 X-ray bursts 
detected from MX 0836-42 using PCA/RXTE data.}
\end{figure}

\begin{table}
\caption{Observations of MX 0836-42 by the JEM-X and IBIS telescopes of the INTEGRAL orbiting observatory}
\vspace{-1.0cm}
\begin{center}
\begin{tabular}{ccccc}
\multicolumn{3}{c}{\it }\\
\hline
\hline

{Beginning of}&
{End of}&
{Exposure}&
{Exposure}&
{Flux}\\
{observation,}&
{observation,}&
{IBIS,}&
{JEM-X,}&
{ISGRI/IBIS (20-120 keV),}\\
{UTC}&
{UTC}&
{s}&
{s}&
{mCrab}\\
  
\hline
 $\bf 2003$&$$&$$&$$&$$ \\
\hline

$ 28.03 $ & $ Mar~ 28 $ & $6571$   & $-$      & $53.94\pm2.16$ \\ 
$ 05.04 $  & $ Apr~ 5 $  & $6600$   & $-$      & $59.95\pm1.77$ \\
$ 20.04 $ & $ Apr~ 20 $ & $6600$   & $-$      & $27.85\pm1.76$ \\ 
$ 29.04 $ & $ Apr~ 29 $ & $6713$   & $4513$   & $6.45\pm1.68$  \\ 
$ 14.05 $ & $ May~ 14 $ & $8802$   & $4399$   & $10.60^a$      \\ 
$ 29.05 $ & $ May~ 29 $ & $6600$   & $2200$   & $16.30^a$      \\ 
$ 05.06 $  & $ Jun~ 5 $  & $8917$   & $2201$   & $9.84^a$       \\
$ 12.06 $ & $ Jun~ 15 $ & $202879$ & $93575$  & $1.95^a$       \\ 
$ 16.06 $ & $ Jun~ 18 $ & $142958$ & $94976$  & $3.13\pm0.39$  \\ 
$ 19.06 $ & $ Jun~ 21 $ & $193103$ & $88973$  & $2.24^a$       \\ 
$ 26.06 $ & $ Jun~ 27 $ & $106271$ & $85499$  & $2.66^a$       \\ 
$ 28.06 $ & $ Jun~ 30 $ & $192907$ & $122629$ & $2.05^a$       \\ 
$ 01.07 $  & $ Jul~ 3 $  & $175646$ & $145142$ & $1.95^a$       \\
$ 04.07 $  & $ Jul~ 6 $  & $113051$ & $98261$  & $2.60^a$       \\
$ 27.11 $ & $ Nov~ 29 $ & $206101$ & $54131$  & $80.56\pm0.38$ \\ 
$ 30.11 $ & $ Dec~ 2 $  & $206942$ & $113931$ & $81.12\pm0.32$ \\ 
$ 03.12 $  & $ Dec~ 5 $  & $186043$ & $97123$  & $77.59\pm0.35$ \\
$ 06.12 $  & $ Dec~ 8 $  & $207723$ & $88880$  & $74.39\pm0.33$ \\
$ 09.12 $  & $ Dec~ 11 $ & $193980$ & $98270$  & $69.01\pm0.32$ \\
$ 24.12 $ & $ Dec~ 24 $ & $6714$   & $4514$   & $69.37\pm2.22$ \\ 

\hline
$\bf 2004$&$$&$$&$$&$$ \\
\hline

$ 02.01 $  & $ Jan~ 2 $  & $4463$ & $2263$ & $57.17\pm2.39$  \\
$ 17.01 $ & $ Jan~ 17 $ & $5713$ & $3513$ & $46.81\pm2.13$  \\ 
$ 26.01 $ & $ Jan~ 26 $ & $6602$ & $2201$ & $50.69\pm1.88$  \\ 
$ 07.02 $  & $ Feb~ 7 $  & $6634$ & $-$    & $58.36\pm2.21$  \\
$ 07.03 $  & $ Mar~ 7 $  & $4388$ & $-$    & $61.62\pm2.55$  \\
$ 26.03 $ & $ Mar~ 26 $ & $6626$ & $-$    & $53.14\pm1.88$  \\ 
$ 24.04 $ & $ Apr~ 24 $ & $6712$ & $4512$ & $45.69\pm1.79$  \\ 
$ 16.05 $ & $ May~ 16 $ & $2314$ & $-$    & $46.95\pm3.83$  \\ 
$ 06.06 $  & $ Jun~ 6 $  & $4400$ & $-$    & $19.78\pm2.34$  \\
$ 11.06 $ & $ Jun~ 11 $ & $6602$ & $2201$ & $11.51\pm1.82$  \\ 
$ 30.06 $ & $ Jun~ 30 $ & $8123$ & $2200$ & $4.62^a$        \\ 
$ 11.07 $ & $ Jul~ 11 $ & $6602$ & $2200$ & $5.17^a$        \\ 
$ 23.07 $ & $ Jul~ 23 $ & $4401$ & $-$    & $8.49^a$        \\ 
$ 21.08 $ & $ Aug~ 21 $ & $5470$ & $-$    & $7.95^a$        \\ 
$ 28.08 $ & $ Aug~ 28 $ & $2221$ & $-$    & $10.83^a$       \\ 
$ 13.09 $ & $ Sep~ 13 $ & $8064$ & $2171$ & $5.14^a$        \\ 
\hline
\hline

\end{tabular} \\

\end{center}
Note. Exposure is the total duration of all pointings during which the source was resolved.\\
$^a$ - 3 $\sigma$ is an upper limit on the energy flux from the source.\\
\end{table}

\begin{table}
\caption{Observations of MX 0836-42 by the PCA and HEXTE instruments of the RXTE orbiting observatory}
\begin{center}
\begin{tabular}{cccccc}

\multicolumn{3}{c}{\it }\\
\hline
\hline

{Date,}&
{Exposure$^a$}&
{Exposure$^a$}&
{$\dot{M}^b \times 10^{-9}$,}&
{$\tau_{1}^c$,}&
{$\tau_{2}^d$,}\\
{UTC}&
{PCA}&
{HEXTE}&
{$M_{_\odot} / yr$}&
{s}&
{s}\\
  
\hline
 
$\bf 2003$&$$&$$&$$&$$&$$  \\
\hline

$ 23.01 $   & $11296$   & $5539$ & $2.08\pm0.03$ &$3230$ &$9680$    \\
$ 24.01 $   & $ 5088$   & $2268$ & $2.09\pm0.04$ &$3180$ &$9810$    \\
$ 25.01 $   & $10592$   & $5689$ & $1.82\pm0.03$ &$3640$ &$10930$   \\
$ 26.01 $   & $14144$   & $7506$ & $1.75\pm0.02$ &$3810$ &$11420$   \\
$ 27.01 $   & $ 9504$   & $5189$ & $1.65\pm0.03$ &$4020$ &$12060$   \\
$ 28_1 .01$ & $ 1872$   & $721 $ & $2.08\pm0.12$ &$3200$ &$9580$    \\
$ 28_2 .01$ & $  864$   & $294 $ & $1.18\pm0.08$ &$5620$ &$16870$   \\
$ 29_1 .01$ & $ 1744$   & $602 $ & $0.88\pm0.12$ &$7560$ &$22680$   \\
$ 29_2 .01$ & $  944$   & $393 $ & $1.31\pm0.05$ &$5070$ &$15200$   \\
$ 30_1 .01$ & $  896$   & $367 $ & $1.74\pm0.05$ &$3820$ &$11450$   \\
$ 30_2 .01$ & $  736$   & $327 $ & $1.23\pm0.05$ &$5410$ &$16220$   \\
$ 30_3 .01$ & $  720$   & $318 $ & $1.74\pm0.06$ &$3820$ &$11460$   \\
$ 31_1 .01$ & $ 5520$   & $2253$ & $1.28\pm0.05$ &$5170$ &$15510$   \\
$ 31_2 .01$ & $ 4032$   & $2128$ & $1.05\pm0.04$ &$6340$ &$19020$   \\
$ 01.02 $   & $ 8992$   & $5085$ & $1.48\pm0.07$ &$4710$ &$13460$   \\
$ 02.02 $   & $ 2944$   & $1413$ & $1.04\pm0.06$ &$6400$ &$19200$   \\
$ 03.02 $   & $ 2464$   & $1007$ & $1.39\pm0.06$ &$4780$ &$14340$   \\
$ 13.03 $   & $ 4672$   & $2211$ & $1.18\pm0.04$ &$5620$ &$16870$   \\
$ 14.03 $   & $10496$   & $5029$ & $1.22\pm0.02$ &$5300$ &$15900$   \\
$ 17.03 $   & $ 3360$   & $1627$ & $1.23\pm0.05$ &$5400$ &$16200$   \\
$ 18.03 $   & $ 3392$   & $1583$ & $1.12\pm0.05$ &$5930$ &$17190$   \\
$ 20.03 $   & $ 3472$   & $1652$ & $1.19\pm0.05$ &$5590$ &$16750$   \\

\hline
$\bf 2004$&$$&$$&$$&$$&$$  \\
\hline

$ 18.01 $   & $ 1120$   & $464$ & $0.90\pm0.47$ &$4990$ &$14960$   \\
$ 23.01 $   & $ 1216$   & $520$ & $0.99\pm0.03$ &$5040$ &$15110$   \\
$ 26.01 $   & $ 1552$   & $723$ & $0.78\pm0.02$ &$6390$ &$19160$   \\

\hline
\hline
\\
\end{tabular} 
\end{center}

Note. The subscript in the dates indicate the pointing number during the corresponding day.\\

$^a$ - The total exposure time in seconds.\\
$^b$ - The accretion rate.\\
$^c$ - The burst recurrence period calculated in the case where 1/3 of the neutron star surface is covered with accreted matter. \\
$^d$ - The burst recurrence period calculated in the case where the neutron star surface is completely covered with accreted matter. \\

\end{table}

\begin{table}
\caption{Results of fitting the RXTE (PCA+HEXTE) spectra of MX 0836-42 by a power law with a high-energy exponential 
cutoff with allowance made for the interstellar absorption and the iron emission line at $E_{Fe}=6.4$ keV 
and the reflection of emission from the accretion disk}
\begin{center}
\begin{tabular}{ccccccc}

\multicolumn{3}{c}{\it }\\
\hline
\hline

{Date,}&
{$\alpha^a$}&
{$E_{_{cut}}^b$,}&
{Flux$^c$,}&
{$N_{_H}^d$,}&
{$K^e$}&
{${\chi^2(N)}^f$}\\
{UTC}&
{}&
{keV}&
{erg/cm$^2$/s}&
{$10^{22}$ at./cm$^{2}$}&
{}&
{}\\
  
\hline
 
$\bf 2003$&$$&$$&$$&$$&$$&$$  \\
\hline
$  23.01 $ & $1.45\pm0.04$ & $53.9\pm4.9$ & $3.18\pm0.05$ & $2.71\pm0.08$ & $0.75$ & $1.27(60)$\\
$  24.01 $ & $1.44\pm0.04$ & $45.9\pm4.6$ & $3.20\pm0.06$ & $2.73\pm0.09$ & $1.00$ & $0.87(66)$\\
$  25.01 $ & $1.47\pm0.03$ & $53.3\pm3.8$ & $2.79\pm0.04$ & $3.07\pm0.06$ & $0.74$ & $1.16(63)$\\
$  26.01 $ & $1.50\pm0.09$ & $67.4\pm3.8$ & $2.67\pm0.04$ & $3.31\pm0.07$ & $0.79$ & $2.08(50)$\\
$  27.01 $ & $1.47\pm0.03$ & $57.9\pm3.1$ & $2.53\pm0.04$ & $3.12\pm0.08$ & $0.76$ & $1.09(59)$\\
$  28_1 .01$ & $1.38\pm0.07$ & $50.6\pm10.1$ & $3.18\pm0.20$ & $2.71\pm0.28$ & $1.03$ & $0.87(51)$\\
$  28_2 .01$ & $1.43\pm0.08$ & $38.2\pm7.7$ & $1.81\pm0.11$ & $4.25\pm0.26$ & $0.57$ & $0.97(52)$\\
$  29_1 .01$ & $0.81\pm0.16$ & $29.4\pm4.6$ & $1.34\pm0.20$ & $6.97\pm0.79$ & $0.77$ & $1.99(52)$\\
$  29_2 .01$ & $1.45\pm0.10$ & $43.7\pm14.7$ & $2.01\pm0.15$ & $2.80\pm0.29$ & $0.93$ & $0.88(44)$\\
$  30_1 .01$ & $1.41\pm0.08$ & $45.8\pm8.2$ & $2.66\pm0.17$ & $3.56\pm0.28$ & $0.97$ & $0.91(52)$\\
$  30_2 .01$ & $1.52\pm0.10$ & $50.0(fixed)$ & $1.88\pm0.11$ & $2.94\pm0.30$ & $1.12$ & $1.05(53)$\\
$  30_3 .01$ & $1.40\pm0.09$ & $56.1\pm14.4$ & $2.66\pm0.19$ & $3.24\pm0.32$ & $0.98$ & $0.93(49)$\\
$  31_1 .01$ & $1.44\pm0.04$ & $50.3\pm5.3$ & $1.95\pm0.07$ & $3.20\pm0.11$ & $0.75$ & $1.25(52)$\\
$  31_2 .01$ & $1.32\pm0.03$ & $43.5\pm4.1$ & $1.61\pm0.06$ & $6.18\pm0.19$ & $0.70$ & $1.88(52)$\\
$  01.02 $ & $1.44\pm0.02$ & $61.0\pm3.8$ & $2.27\pm0.03$ & $3.35\pm0.07$ & $0.54$ & $1.27(66)$\\
$  02.02 $ & $1.16\pm0.08$ & $35.3\pm3.4$ & $1.59\pm0.10$ & $9.86\pm0.32$ & $0.87$ & $1.29(50)$\\
$  03.02 $ & $1.46\pm0.05$ & $53.4\pm7.9$ & $2.13\pm0.09$ & $3.23\pm0.19$ & $0.90$ & $1.03(52)$\\
$  13.03 $ & $1.51\pm0.05$ & $65.8\pm8.3$ & $1.80\pm0.06$ & $2.27\pm0.16$ & $0.71$ & $1.27(55)$\\
$  14.03 $ & $1.54\pm0.04$ & $64.9\pm5.5$ & $1.87\pm0.04$ & $2.73\pm0.09$ & $0.76$ & $1.08(50)$\\
$  17.03 $ & $1.54\pm0.06$ & $76.2\pm12.6$ & $1.88\pm0.08$ & $2.33\pm0.20$ & $0.86$ & $0.88(55)$\\
$  18.03 $ & $1.51\pm0.06$ & $67.5\pm12.4$ & $1.71\pm0.08$ & $2.46\pm0.21$ & $0.81$ & $0.98(52)$\\
$  20.03 $ & $1.58\pm0.06$ & $71.4\pm14.1$ & $1.82\pm0.07$ & $2.58\pm0.17$ & $0.92$ & $0.87(48)$\\

\hline
$\bf 2004$&$$&$$&$$&$$&$$&$$\\
\hline

$  18.01 $ & $1.49\pm0.07$ & $50.0(fixed)$ & $1.37\pm0.08$ & $1.83\pm0.28$ & $1.01$ & $0.96(54)$\\
$  23.01 $ & $1.54\pm0.07$ & $50.0(fixed)$ & $1.51\pm0.08$ & $1.88\pm0.25$ & $0.98$ & $0.93(54)$\\
$  26.01 $ & $1.59\pm0.07$ & $53.5\pm15.2$ & $1.19\pm0.08$ & $1.90\pm0.26$ & $0.65$ & $1.27(53)$\\

\hline
\hline
\\
\end{tabular} 
\end{center}
$^a$ - The photon index. \\
$^b$ - The exponential cutoff energy. \\
$^c$ - The persistent (3-120 keV) flux from the source ($\times 10^{-9}$~erg ~cm $^{-2}$ ~s $^{-1}$) \\
$^d$ - The hydrogen column density obtained when fitting the spectrum. \\
$^e$ - The scaling factor of the HEXTE spectrum. \\
$^f$ - The $\chi ^2$ value of the best fit to the spectrum normalized to the number of degrees of freedom N. \\

\end{table}

\begin{table}
\caption{X-ray bursts detected from MX 0836-42 by the JEM-X instrument of the INTEGRAL orbiting observatory}
\begin{center}
\begin{tabular}{cccccc}

\multicolumn{3}{c}{\it }\\
\hline
\hline

{Date$^a$,}&
{${F_{m}}^b$,}&
{Date$^a$,}&
{${F_{m}}^b$,}&
{Date$^a$,}&
{${F_{m}}^b$,}\\
{MJD}&
{Crab}&
{MJD}&
{Crab}&
{MJD}&
{Crab}\\

\hline
\hline
  
 $52971.954857$ & $0.83\pm0.20$ & $52975.559510$ & $0.75\pm0.20$ & $52980.954001$ & $0.72\pm0.18$  \\
 $52972.054063$ & $0.62\pm0.15$ & $52977.293263$ & $0.85\pm0.21$ & $52981.048922$ & $0.74\pm0.19$  \\
 $52972.158769$ & $0.93\pm0.22$ & $52977.465705$ & $0.69\pm0.17$ & $52981.146508$ & $0.63\pm0.16$  \\
 $52973.403827$ & $0.87\pm0.21$ & $52978.798908$ & $0.85\pm0.26$ & $52981.242242$ & $0.56\pm0.17$  \\
 $52974.639704$ & $0.76\pm0.21$ & $52979.840899$ & $1.12\pm0.24$ & $52982.428908$ & $0.51\pm0.16$  \\
 $52974.733813$ & $0.69\pm0.19$ & $52979.930894$ & $0.77\pm0.17$ & $52983.507589$ & $0.57\pm0.18$  \\
 $52975.372256$ & $0.72\pm0.24$ & $52980.037844$ & $0.84\pm0.21$ & $52983.710369$ & $0.70\pm0.17$  \\
 $52975.386177$ & $0.55\pm0.23$ & $52980.124557$ & $0.73\pm0.18$ & $52984.442369$ & $0.85\pm0.25$  \\
  
\hline
\hline
\\
\end{tabular} 
\end{center}
$^{a}$ - The time the flux reaches its maximum. \\
$^{b}$ - The maximum burst flux averaged over 1 s. \\

\end{table}

\begin{table}
\caption{X-ray bursts detected from MX 0836-42 by PCA aboard the RXTE observatory}
\begin{center}
\begin{tabular}{ccccccccc}

\multicolumn{3}{c}{\it }\\
\hline
\hline

{Date,}&
{${T}^a$,}&
{${T_{exp}}^b,$}&
{${T_{eff}}^c,$}&
{${F_{max}}^d,$}&
{${F_{_p}}^e$,}&
{${E_{_b}}^f$,}&
{${\tau_{_R}}^g$,}\\
{UTC}&
{MJD}&
{s}&
{s}&
{erg cm $^{-2}$ s $^{-1}$}&
{erg cm $^{-2}$ s $^{-1}$}&
{erg}&
{s}\\
  
\hline
 
$\bf 2003$&$$&$$&$$&$$&$$&$$&$$  \\
\hline

$ 23.01   $ & $52662.593715$ & $8.3$  & $16.3$ & $1.63\pm0.10$ & $3.18\pm0.05$ & $2.04\pm0.45$ & $11490$  \\
$ 24.01   $ & $52663.305984$ & $13.0$ & $25.2$ & $1.23\pm0.07$ & $3.26\pm0.06$ & $2.37\pm0.30$ & $13540$  \\
$ 25_1 .01$ & $52664.444097$ & $17.3$ & $21.2$ & $1.08\pm0.06$ & $2.56\pm0.13$ & $1.75\pm0.24$ & $12500$  \\
$ 25_2 .01$ & $52664.630428$ & $19.4$ & $19.1$ & $1.01\pm0.06$ & $2.96\pm0.07$ & $1.48\pm0.36$ & $9140$  \\
$ 26.01   $ & $52665.702338$ & $15.9$ & $15.4$ & $1.14\pm0.06$ & $2.67\pm0.04$ & $1.34\pm0.19$ & $9180$  \\
$ 27.01   $ & $52666.560185$ & $20.4$ & $26.5$ & $0.98\pm0.07$ & $2.53\pm0.04$ & $1.99\pm0.27$ & $14380$  \\
$ 28.01   $ & $52667.668495$ & $14.7$ & $21.3$ & $0.95\pm0.07$ & $1.81\pm0.11$ & $1.55\pm0.25$ & $15660$  \\
$ 29.01   $ & $52668.191574$ & $12.4$ & $16.7$ & $0.82\pm0.07$ & $1.34\pm0.20$ & $1.05\pm0.20$ & $14320$  \\
$ 31_1 .01$ & $52670.553704$ & $12.6$ & $18.3$ & $1.18\pm0.07$ & $1.95\pm0.05$ & $1.65\pm0.24$ & $15470$  \\
$ 31_2 .01$ & $52670.637095$ & $10.9$ & $12.3$ & $0.66\pm0.05$ & $1.61\pm0.05$ & $0.62\pm0.14$ & $7205$  \\
$ 02.02   $ & $52672.624456$ & $16.9$ & $28.6$ & $0.89\pm0.06$ & $1.59\pm0.10$ & $1.95\pm0.21$ & $22430$  \\
$ 14.03   $ & $52712.408229$ & $16.0$ & $17.6$ & $1.36\pm0.08$ & $1.87\pm0.04$ & $1.83\pm0.29$ & $17890$  \\
$ 17.03   $ & $52715.453229$ & $16.6$ & $16.0$ & $1.12\pm0.09$ & $1.88\pm0.08$ & $1.37\pm0.21$ & $13320$  \\
$ 20.03   $ & $52718.536470$ & $16.3$ & $25.6$ & $1.03\pm0.07$ & $1.82\pm0.07$ & $2.02\pm0.39$ & $20290$  \\

\hline
$\bf 2004$&$$&$$&$$&$$&$$&$$&$$ \\
\hline

$ 18.01 $ & $53022.663252$ & $16.8$ & $18.7$ & $0.92\pm0.06$ & $1.37\pm0.08$ & $1.32\pm0.24$ & $17620$  \\

\hline
\hline
\\
\end{tabular} 
\end{center}

Note. The subscript in the dates indicates the burst number during the corresponding day. \\

$^a$ - The burst onset time. \\
$^b$ - The exponential burst decay time. \\
$^c$ - The effective burst duration. \\
$^d$ - The 3-20 keV maximum burst flux ($\times 10^{-8}$)\\
$^e$ - The persistent 3-120 keV flux ($\times 10^{-9}$)\\
$^f$ - The energy released during the burst ($\times (D/8$ kpc $)^2\times 10^{39}$)\\
$^g$ - The burst recurrence period (at $\alpha=140$)\\

\end{table}

\end{document}